\documentclass[showpacs,aps,graphicx,twocolumn]{revtex4}
\usepackage{amssymb}
\usepackage{graphicx}

\begin{document}

\title{ Complete hyperentangled-Bell-state analysis for quantum communication\footnote{Published in Phys. Rev. A \textbf{82}, 032318 (2010)}}

\author{Yu-Bo Sheng$^{1,2,3}$, Fu-Guo Deng$^{2}$,   and Gui Lu Long$^{1,4,5}$\footnote{Corresponding author:
gllong@tsinghua.edu.cn}}
\address{$^1$ Department of Physics, Tsinghua University,
Beijing 100084  China\\
$^2$ Department of Physics,  Beijing Normal University, Beijing 100875, China\\
$^3$ College of Nuclear Science and Technology, Beijing Normal
University, Beijing 100875, China\\
 $^4$ Center for Atomic and
Molecular NanoSciences, Tsinghua University,
Beijing 100084, China\\
$^5$ Key Laboratory For Quantum Information and Measurements,
Beijing 100084,  China }

\date{\today }

\begin{abstract}
It is impossible to unambiguously distinguish the four Bell states
in polarization, resorting to linear optical elements only.
Recently, the hyperentangled Bell state, the simultaneous
entanglement in more than one degree of freedom, has been used to
assist in the complete Bell-state analysis of the four Bell states.
However, if the additional degree of freedom is qubitlike, one can
only distinguish 7 from the group of 16 states. Here we present a
way to distinguish the hyperentangled Bell states completely with
the help of cross-Kerr nonlinearity. Also, we discuss its
application in the quantum teleportation of a particle in an unknown
state in two different degrees of freedom and in the entanglement
swapping of hyperentangled states. These applications will increase
the channel capacity of long-distance quantum communication.
\end{abstract}

 \maketitle

\section{introduction}

Quantum teleportation \cite{teleportation,cteleportation}, quantum
dense coding  \cite{densecoding}, quantum superdense coding
\cite{supdensecoding}, and some important quantum cryptographic
schemes \cite{Ekert91,BBM92,longqkd,CORE}, which are the completion
of most fundamental quantum communication processes involving
bipartite entanglement, need the complete and deterministic analysis
of the Bell states. For instance, in the well-known teleportation
protocol \cite{teleportation}, an unknown qubit state may be
teleproted between two parties over a long distance as long as each
of them possesses one particle in an entangled photon pair and then
the sender, Alice, makes an appropriate joint measurement on her
particle in an unknown state  and one of the pair shared with the
receiver Bob. Unfortunately, with only linear optical elements, a
complete Bell-state analysis (BSA) is impossible and one can only
get the optimal success probability of $\frac{1}{2}$
 \cite{vaidman,bellmeasurement1,bellmeasurement2}. In the optimal optical BSA
scheme realized experimentally, its success probability is only 50\%
\cite{mattle,hou,ursin}.

Hyperentanglement  \cite{hyper7,hyper8,hyper9}, the simultaneous
entanglement in more than one degree of freedom, such as
polarization-momentum, polarization-time-bin, and polarization- and
spatial-modes-energy-time, can be used to assist the complete
Bell-state discrimination
\cite{hyper1,hyper2,hyper3,hyper4,hyper5,hyper6}. For instance,
Kwait and Weinfurter  \cite{hyper1} first discussed the way to
distinguish the four orthogonal Bell states with both momentum
entanglements and polarization entanglements. In 2003, Walborn
\emph{et al.} \cite{hyper2} proposed a simple scheme for complete
Bell-state measurement for photons using hyperentangled states. In
their protocol, they performed the polarization BSA by using
momentum entanglement as an ancilla. They also showed that the
polarization states can be used to distinguish the momentum Bell
states. The experiments of a complete BSA have also been reported
with polarization-time-bin hyperentanglement \cite{hyper3} and
polarization-momentum hyperentanglement \cite{hyper4}. In 2008,
Kwait \emph{et al.} beat the channel capacity limit for linear
photonic superdense coding with polarization-orbital angular
momentum hyperentanglement \cite{hyper5}. In essence, a BSA in the
polarization degree of freedom with hyperentanglement  works in a
larger Hilbert space by introducing other degrees of freedom.

If  we consider a large Hilbert space with an additional degree of
freedom, the case is quite different.  For a quantum system in a
hyerentangled state in two degrees of freedom, it has 16 orthogonal
Bell states. With  linear optics only, one cannot distinguish them
completely. For instance, the momentum entanglement can be used to
distinguish the four polarization Bell states completely, but the
momentum entanglement itself cannot be distinguished well. In 2007,
Wei \emph{et al.}  \cite{hyper6} pointed out that one can only
distinguish 7 states in the group of 16  orthogonal Bell states.
 If we consider each photon in $n$ qubit-like degrees of
freedom, there are, in total, $4^{n}$ Bell-like states for each
two-photon quantum system. Wei \emph{et al.}  \cite{hyper6} also
extended their result and showed that  the upper bound of the
maximal number of mutually distinguishable Bell-like states is
$2^{n+1}-1$, as  is true for $n=1$ and $n=2$.

In this article, we  present a complete hyperentanglement BSA (HBSA)
with cross-Kerr nonlinearity. The hyperentangled state we discussed
comprises the entanglement with two different kinds of degrees of
freedom, that is, polarization and spatial modes. This HBSA scheme
can be divided into two steps. The first step is used to distinguish
the Bell states in spatial modes but not destroy the two photons.
With this step, one can not only read the information about the Bell
states in spatial modes, but also confirm the outports of the two
photons, which provides  useful information for the BSA in
polarization. This task should resort to quantum nondemolition
detectors (QNDs). In the second step, one can first divide the four
Bell states in polarization into two groups according to their
parities, that is, the even-parity states and the odd-parity states
with QND, and then distinguish the two Bell states in each group
with two Hadamard operations and two parity-check measurements. This
HBSA protocol can be used in quantum teleportation of a
single-particle system in an unknown state in the polarization and
spatial-mode degrees of freedom. Also, it has a good application in
hyperentanglement swapping of a hyerentangled state in two different
degrees of freedom. These applications will increase the channel
capacity in long-distance quantum communication.

This article is organized as follows: In Sec.II, we  explain the
principle of the present HBSA scheme. It is divided into two steps.
One is used to distinguish the four Bell states in the spatial-mode
degree of freedom with QNDs, which is discussed in Sec.II A. The
other is used to analyze the four Bell states in the polarization
degree of freedom, including a parity-check measurement with QND and
another one with linear optical elements, shown in Sec.II B. In
Sec.III, we discuss the applications of the present HBSA scheme in
long-distance quantum communication; that is, a hyperentanglement
teleportation scheme and a hyperentanglement-swapping scheme are
discussed in Sec.III A and B, respectively. A discussion and a
summary are given in Sec.IV.

\section{Complete hyperentanglement stateBSA}

\subsection{ HBSA protocol for Bell  states in spatial modes }

The previous works indicated  that  QNDs with a cross-Kerr medium
and a coherent state can be used for operating the controlled-not
(CNOT) gate \cite{QND1,lin1} and single-photon logic gates with
minimal sources \cite{lin2}, entanglement purification and
concentration \cite{shengpra,shengpra2}; generating high-quality
entanglement \cite{he1,he2}, and qubits \cite{qubit1,qubit2,qubit3};
and analyzing the Bell states with the polarization degree of
freedom \cite{QND2}. In general, cross-Kerr nonlinearities can be
described with the Hamiltonian as \cite{QND1,QND2}
\begin{eqnarray}
H_{ck}=\hbar\chi a^{\dagger}_{s}a_{s}a^{\dagger}_{p}a_{p}.
\end{eqnarray}
Here $a_{s}$ and $a_{p}$ are the annihilation operations, and
$a^{\dagger}_{s}$ and $a^{\dagger}_{p}$ denote the creation
operations.  $\hbar \chi $ is the coupling strength of the
nonlinearity and it is decided by the property of the nonlinear
material. If we consider a single-photon state
$|\varphi\rangle=a|0\rangle+b|1\rangle$ and a coherent state
$|\alpha\rangle$, the cross-Kerr interaction causes the combined
system composed of a single photon and a coherent state  evolve as
\begin{eqnarray}
U_{ck}|\varphi\rangle|\alpha\rangle &=& e^{iH_{QND}t/\hbar}(a|0\rangle+b|1\rangle)|\alpha\rangle \nonumber\\
&=& a|0\rangle\alpha\rangle+b|1\rangle|\alpha e^{i\theta}\rangle,
\end{eqnarray}
where  $|0\rangle$ and $|1\rangle$ are the Fock states which means
the states contain 0 and 1 photon, respectively.  Here the phase
shift $\theta=\chi t$ and $t$ is the interaction time which is
directly proportional to the number of photons with the
single-photon state being unaffected. In 2005, Barratt \emph{et al.}
performed a polarization BSA with QND \cite{QND2}. They first used
the QND to construct the parity-check measurement.

Now we will work with a hyperentangled two-photon state with the
form
\begin{eqnarray}
|\Phi^{+}_{ab}\rangle_{PS}=\frac{1}{2}(|HH\rangle+|VV\rangle)_{ab}\otimes(|a1b1\rangle+|a2b2\rangle)_{ab}.\label{hyperentanglement}
\end{eqnarray}
$|H\rangle$ and $|V\rangle$ are the horizontal and the vertical
polarizations, respectively. The subscripts $a$ and $b$ represent
the two photons in the hyperentangled state. The subscript $P$
denotes the polarization degree of freedom and $S$ is the
spatial-mode degree of freedom. $a1(b1)$ and $a2(b2)$ are the
different spatial modes for photon $a(b)$, shown in Fig.1. The state
of Eq.(3) can be easily produced with a parametric down-conversion
source. For example, in Ref. \cite{Simon}, a pump pulse comes from
below and traverses a nonlinear $\beta$ barium borate (BBO) crystal,
where it can produce the entangled state into the modes $a1b1$, then
it is reflected and traverses the BBO crystal twice to produce
entangled pairs in the modes $a2b2$. The photon pairs also are
entangled in the polarization degree of freedom. That is, each
photon pair is in the hyperentangled state
$|\Phi^{+}_{ab}\rangle_{PS}$.

We denote four Bell states  in the polarization degree of freedom as
\begin{eqnarray}
|\phi^{\pm}\rangle_{P}=\frac{1}{\sqrt{2}}(|HH\rangle\pm|VV\rangle),\nonumber\\
|\psi^{\pm}\rangle_{P}=\frac{1}{\sqrt{2}}(|HV\rangle\pm|VH\rangle),\label{polarization}
\end{eqnarray}
and four Bell states in the spatial-mode  degree of freedom as
\begin{eqnarray}
|\phi^{\pm}\rangle_{S}=\frac{1}{\sqrt{2}}(|a1b1\rangle\pm|a2b2\rangle),\nonumber\\
|\psi^{\pm}\rangle_{S}=\frac{1}{\sqrt{2}}(|a1b2\rangle\pm|a2b1\rangle).\label{spatial}
\end{eqnarray}
 Sometimes we refer to the states
$|\phi^{\pm}\rangle_{P}$ and $|\phi^{\pm}\rangle_{S}$ as the
even-parity states, and $|\psi^{\pm}\rangle_{P}$ and
$|\psi^{\pm}\rangle_{S}$ as the odd-parity states.

An appealing advantage of the hyperentangled product states with the
form of Eq.(\ref{hyperentanglement}) is that state in the two
degrees of freedom can be well operated independently. For example,
if we manipulate the polarization Bell states using some local
operations, such as a polarization beam splitter (PBS) or a
half-wave plate (HWP), we will leave the spatial-mode entangled
state unchanged. Meanwhile, if we operate the spatial-mode Bell
states, the polarization Bell states remain unchanged too. This
feature provides us an effective way of achieving HBSA. That is, one
can perform the polarization BSA and spatial-mode Bell-state
analysis independently. The spatial-mode entanglement no longer acts
as an ancilla for the polarization BSA. Therefore, a HBSA can be
divided into two steps. The first step is used for the spatial-mode
BSA and the second is for polarization. The precondition for this
scheme being well realized is that of   nondestructive measurement,
as one can not operate polarization BSA any more if the photons are
detected and destroyed. Fortunately, cross-Kerr nonlinearities
provide us a powerful tool for accomplishing QNDs.

\begin{figure}[!ht]
\begin{center}
\includegraphics[width=6.8cm,angle=0]{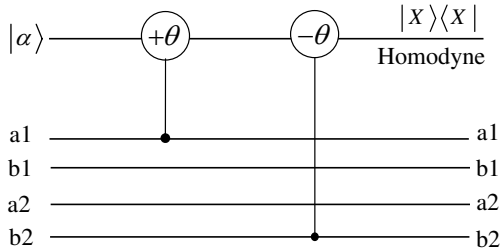}
\caption{Schematic diagram of the present HBSA protocol for
spatial-mode entangled Bell states. This QND is used to distinguish
the even-parity states $\{\vert \phi^\pm\rangle_S\}$ from the
odd-parity states $\{\vert \psi^\pm\rangle_S\}$. The action of first
cross-Kerr nonlinearity puts a phase shift $\theta$ on the coherent
probe beam if a photon appears in the mode coupled. The second
cross-Kerr nonlinearity puts a phase shift with $-\theta$. After the
nonlinear interactions, the probe beam picks up the phase shift
$\theta$ or $-\theta$  if the state is $|a1b1\rangle$ or
$|a2b2\rangle$, respectively. Otherwise, the states $|a1b2\rangle$
and $|a2b1\rangle$ pick up no phase shifts.}
\end{center}
\end{figure}

The principle of the present HBSA protocol for spatial-mode
entangled Bell states is shown in Fig.1 and Fig.2. With the first
QND shown in Fig.1, the state $|\phi^{\pm}\rangle_{S}$ with the
coherent state $|\alpha\rangle$ evolves as
\begin{eqnarray}
|\phi^{\pm}\rangle_{S}|\alpha\rangle &=& \frac{1}{\sqrt{2}}(|a1b1\rangle\pm|a2b2\rangle)|\alpha\rangle\nonumber\\
&\rightarrow&\frac{1}{\sqrt{2}}(|a1b1\rangle|\alpha
e^{i\theta}\rangle\pm|a2b2\rangle|\alpha e^{-i\theta}\rangle),
\end{eqnarray}
but the state $|\psi^{\pm}\rangle_{S}$ with the coherent state
$|\alpha\rangle$ evolves as
\begin{eqnarray}
|\psi^{\pm}\rangle_{S}|\alpha\rangle &=& \frac{1}{\sqrt{2}}(|a1b2\rangle\pm|a2b1\rangle)|\alpha\rangle\nonumber\\
&\rightarrow&\frac{1}{\sqrt{2}}(|a1b2\rangle|\alpha
\rangle\pm|a2b1\rangle|\alpha\rangle)
\nonumber\\
&=& |\psi^{\pm}\rangle_{S}|\alpha \rangle.
\end{eqnarray}
One can observe immediately that the states $|a1b2\rangle$ and
$|a2b1\rangle$ pick up no phase shifts and maintain the coherence
with respect to each other. However, the states $|a1b1\rangle$ and
$|a2b2\rangle$ pick up the phase shifts $\theta$ and  $-\theta$,
respectively. If we choose an X quadrature measurement on the
coherent beam,  with which the states $|\alpha e^{-i\theta}\rangle$
and $|\alpha e^{i\theta}\rangle$ cannot be distinguished
\cite{QND1}, we can distinguish $|\phi^{\pm}\rangle_{S}$ and
$|\psi^{\pm}\rangle_{S}$ with different phase shifts by
homodyne-heterodyne measurements. This QND detector is a
parity-checking device which can distinguish the even-parity states
$|\phi^{\pm}\rangle_{S}$ from the odd-parity states
$|\psi^{\pm}\rangle_{S}$.

\begin{figure}[!ht]
\begin{center}
\includegraphics[width=8cm,angle=0]{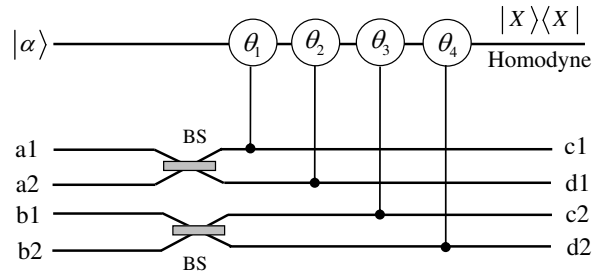}
\caption{The second QND for the BSA in spatial modes. 50:50 BS acts
as   a Hadamard operation. $\theta_i$ ($i=1,2,3,4$) represent four
different cross-Kerr nonlinear media with the phase shifts
$\theta_i$. This QND is used to distinguish the four spatial modes
$c1c2$, $d1d2$, $c1d2$, and $d1c2$ with the phase shifts $\theta_1 +
\theta_3$, $\theta_2 + \theta_4$, $\theta_1 + \theta_4$, and
$\theta_2 + \theta_3$, respectively. When the two photons appear at
the spatial modes $c1c2$ or  $d1d2$ ($c1d2$ or $d1c2$), they are
originally in the Bell state $\vert \phi^+\rangle_s$ ($\vert
\phi^-\rangle_s$) if the phase shift in the first QND is $\theta$ or
$-\theta$. The result is kept for the state $\vert \psi^+\rangle_s$
($\vert \psi^-\rangle_s$) when the phase shift in the first QND is
zero. Completing the BSA with the first and the second QNDs, the
outports of the two photons are determinate, which provides   useful
information for the analysis of the Bell states in the polarization
degree of freedom.}
\end{center}
\end{figure}

With the first QND, the four Bell states in spatial modes are
divided into two groups $\vert \phi^\pm\rangle_S$ and $\vert
\psi^\pm\rangle_S$. The next task of BSA in spatial modes is to
distinguish the different phases in each group. The second QND is
used to distinguish the Bell state with the phase zero and the Bell
state with the phase $\pi$, shown in Fig.2. A 50:50 beam splitter
(BS) can accomplish the following transformation in the spatial
modes:
\begin{eqnarray}
|a1\rangle\rightarrow\frac{1}{\sqrt{2}}(|c1\rangle+|d1\rangle),\nonumber\\
|a2\rangle\rightarrow\frac{1}{\sqrt{2}}(|c1\rangle-|d1\rangle),\nonumber\\
|b1\rangle\rightarrow\frac{1}{\sqrt{2}}(|c2\rangle+|d2\rangle),\nonumber\\
|b2\rangle\rightarrow\frac{1}{\sqrt{2}}(|c2\rangle-|d2\rangle).
\end{eqnarray}
For the group $|\phi^{\pm}\rangle_{S}$,  the state
$\frac{1}{\sqrt{2}}(|a1b1\rangle+|a2b2\rangle)$ will become
$\frac{1}{\sqrt{2}}(|c1c2\rangle+|d1d2\rangle)$ after the BSs, but
$\frac{1}{\sqrt{2}}(|a1b1\rangle-|a2b2\rangle)$ will become
$\frac{1}{\sqrt{2}}(|c1d2\rangle+|d1c2\rangle)$. From Fig.2, one can
see states $c1c2$, $d1d2$, $c1d2$, and $d1c2$ with the phase shifts
$\theta_1 + \theta_3$, $\theta_2 + \theta_4$, $\theta_1 + \theta_4$,
and $\theta_2 + \theta_3$, respectively. With an X quadrature
measurement on the coherent beam, one can read out the information
about the phases in the group $|\phi^{\pm}\rangle_{S}$. That is, the
two entangled photons are originally in the Bell state
$|\phi^{+}\rangle_{S}$ when they appear at  outport $c1c2$ or
$d1d2$; otherwise, they are originally in the state
$|\phi^{-}\rangle_{S}$.

The states $|\psi^{\pm}\rangle_{S}$ can also be distinguished with
the same method discussed previously. That is, the state $\vert
\psi^+\rangle_S$ will become
$\frac{1}{\sqrt{2}}(|c1c2\rangle-|d1d2\rangle)$ and the state $\vert
\psi^-\rangle_S$ will become
$\frac{1}{\sqrt{2}}(|c1d2\rangle-|d1c2\rangle)$ after the two BSs.
If the two entangled photons appear at the outports $c1c2$ or
$d1d2$, they are originally in the Bell state $\vert
\psi^+\rangle_S$; otherwise, they are originally in the Bell state
$\vert \psi^-\rangle_S$.

From the preceding analysis, one can see that the role of the two
QNDs is to accomplish the task of parity check. The first QND can
distinguish the two even-parity states in spatial modes $\vert
\phi^{\pm}\rangle_S$ from the two odd-parity states
$\psi^{\pm}\rangle_S$. With two BSs, the two states with two
different relative phases are transformed into two states with
different parities. With the second QND, one can in principle
distinguish the four Bell states in spatial modes, without
destroying the two photons. Moreover, the X quadrature measurement
on the coherent beam will give useful information about the outports
of the two entangled photons, which will make the HBSA for the four
Bell states in polarization more convenient.

\subsection{HBSA protocol for Bell states in polarization}

Now let us move our attention to distinguish the four Bell states
$\vert \phi^\pm\rangle_P$ and  $\vert \psi^\pm\rangle_P$ in
polarization. From  the preceding analysis, the spatial-mode
entangled states have been deterministically discriminated. Suppose
that one gets the entangled state in spatial modes
$|\phi^{+}\rangle_{S}=\frac{1}{\sqrt{2}}(|c1c2\rangle +
|d1d2\rangle)$ with QNDs shown in Fig.1 and Fig.2. In fact, we need
only to discuss the case that the photon pair is in the spatial
modes $c1c2$ (the spatial modes $a1b1$ in Fig.3), as the case in the
modes $d1d2$ is similar to it. The analysis of the four Bell states
in polarization  in the present protocol is similar to that in Ref.
\cite{QND2}.

\begin{figure}[!ht]
\begin{center}
\includegraphics[width=6cm,angle=0]{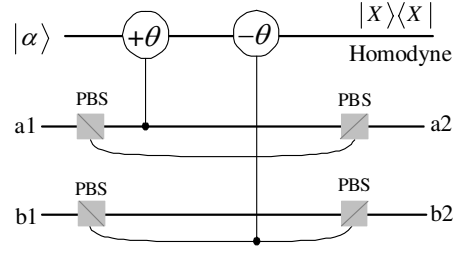}
\caption{Schematic diagram of the present HBSA protocol for Bell
states in polarization. The state $|HH\rangle$ picks up the phase
shift $\theta$ and $|VV\rangle$ picks up the phase shift $-\theta$.
The states $|HV\rangle$ and $|VH\rangle$ pick up no phase shifts.
That is, this setup is a parity-check device for the four Bell
states in polarization.}
\end{center}
\end{figure}

The setup for the discrimination of the four Bell states in
polarization is shown in Fig.3. Cross-Kerr nonlinearities with the
combined system $|\phi^{\pm}\rangle_P|\alpha\rangle$ will evolve as
\begin{eqnarray}
|\phi^{\pm}\rangle_P|\alpha\rangle &=& \frac{1}{\sqrt{2}}(|HH\rangle\pm|VV\rangle)|\alpha\rangle\nonumber\\
&\rightarrow& \frac{1}{\sqrt{2}}(|HH\rangle
|\alpha e^{i\theta}\rangle \pm |VV\rangle|\alpha e^{-i\theta}\rangle).
\end{eqnarray}
$|\psi^{\pm}\rangle_P|\alpha\rangle$ will evolve as
\begin{eqnarray}
|\psi^{\pm}\rangle_P|\alpha\rangle &=&
\frac{1}{\sqrt{2}}(|HV\rangle\pm|VH\rangle)|\alpha\rangle \nonumber\\
&\rightarrow& \frac{1}{\sqrt{2}}(|HV\rangle|\alpha\rangle
\pm|VH\rangle|\alpha\rangle)\nonumber\\
&=&|\psi^{\pm}\rangle_P|\alpha\rangle.
\end{eqnarray}
In these evolutions, $|HV\rangle$ and $|VH\rangle$ pick up no phase
shifts and maintain the coherent state with respect to each other,
but $|HH\rangle$ and $|VV\rangle$ pick up the phase shifts $\theta$
and $-\theta$, respectively. Similar to the preceding case for
spatial-mode entangled states, we choose an X quadrature measurement
on the coherent beam to make $|\alpha e^{\pm i\theta}\rangle$  not
be distinguished.

\begin{figure}[!ht]
\begin{center}
\includegraphics[width=5.4cm,angle=0]{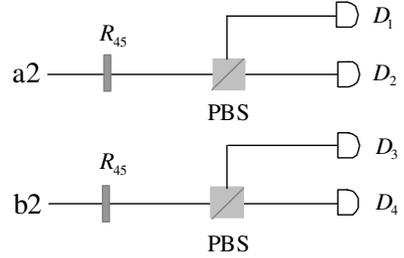}
\caption{The setup for distinguishing the relative phases of the
Bell states in polarization. PBS: polarization beam splitter which
is used to pass through $\vert H\rangle$ polarization photons and
reflect $\vert V\rangle$ polarization photons. The wave plates
$R_{45}$ rotate the horizontal and vertical polarizations by
$45^\circ$, which accomplishes   a Hadamard operation on
polarization. }
\end{center}
\end{figure}

With the QND shown in Fig.3, the four Bell states in polarization
are divided into two groups, that is, the even-parity states $\vert
\phi^{\pm}\rangle_P$ and the odd-parity states $\vert
\psi^{\pm}\rangle_P$. The next step is to distinguish the different
relative phases in each group. This task can be accomplished with
linear optical elements, shown in Fig.4. One can use $\lambda/4$
wave plates $R_{45}$ to rotate the two  photons  $a$ and $b$ by
$45^{\circ}$. The unitary transformation of $45^{\circ}$ rotations
can be described as
\begin{eqnarray}
|H\rangle_{a2} & \rightarrow & \frac{1}{\sqrt{2}}(|H\rangle_{a2}+|V\rangle_{a2}),\nonumber\\
|H\rangle_{b2} & \rightarrow & \frac{1}{\sqrt{2}}(|H\rangle_{b2}+|V\rangle_{b2}),\nonumber\\
|V\rangle_{a2} & \rightarrow & \frac{1}{\sqrt{2}}(|H\rangle_{a2}-|V\rangle_{a2}),\nonumber\\
|V\rangle_{b2} & \rightarrow &
\frac{1}{\sqrt{2}}(|H\rangle_{b2}-|V\rangle_{b2}).
\end{eqnarray}
The state $\vert \phi^+\rangle_P$ ($\vert \psi^-\rangle_P$) is kept
unchanged after the two rotations $R_{45}$ and the two photons will
click the detectors $D_1D_3$ or $D_2D_4$ ($D_1D_4$ or $D_2D_3$). The
state $\vert \phi^-\rangle_P$ ($\vert \psi^+\rangle_P$) will become
$\vert \psi^+\rangle_P$ ($\vert \phi^-\rangle_P$) and the two
photons will click the detectors $D_1D_4$ or $D_2D_3$ ($D_1D_3$ or
$D_2D_4$).

In fact, the BSA in polarization essentially equals  the BSA in
spatial modes. In Fig.3, the polarization entangled states in the
modes $a1b1$ are divided into four different spatial modes by two
PBSs. The QND serves  the same purpose as those in Fig.1 and Fig.2.
The process of the discrimination of the Bell states can be divided
into two steps, with each step being a parity-check measurement.
Thus, the present HBSA protocol can also be served as parity-check
measurements for different degrees of freedom of photons. The
distinct feature of the hyperentangled state is that the different
degrees  of freedom are relatively independent of each other, which
ensures that one can manipulate each degree of freedom
independently. In theory, one can also distinguish the  Bell states
in polarization mode first and then the BSA in spatial mode
modestly. However, this strategy will make the whole discrimination
scheme more complicated as the spatial mode is uncertain, which will
lead us to add more QNDs to accomplish the parity-check
measurements.

\section{Applications of HBSA in quantum communication}

So far, we have described a full HBSA with QNDs. It is interesting
to discuss the applications of HBSA in quantum communication. There
are two unique techniques in long-distance quantum communication,
that is, quantum teleportation and entanglement swapping. The former
can be used to transmit an unknown state of a particle to a remote
point without distributing the particle itself. The latter provides
a good tool for quantum repeaters in long-distance quantum
communication. We discuss the way to teleport an unknown state in
two degrees of freedom of photons and the way to swap two
hyerentangled states in what follows.

\subsection{Teleportation with a hyperentanglement-state channel}

A Bell state shared  enables the teleportation of an unknown
single-qubit state. In a quantum teleportation protocol
\cite{teleportation}, two parties say Alice and Bob in distant
locations are in possession of one photon of a polarization
entangled pair, which is prepared in a Bell state. Alice wants to
teleport another photon to Bob but he does not know any information
about its state, otherwise she only needs to tell Bob to prepare it
with classical communication. For this end, Alice first makes a
Bell-state measurement on her teleported photon and the photon in
the entangled pair shared with Bob and then tells her result to Bob
with classical communication. Finally, Bob can recover the state of
Alice's photon   according to her results with some local unitary
operations. In 1997, the teleportation protocol based on
polarization entanglement was demonstrated experimentally
\cite{teleportationexperiment1} and the teleportation was also
realized by using path-entangled (spatial-mode) photons in 1998
\cite{teleportationexperiment2}.

With linear optics only, the complete BSA of a two-photon
polarization state alone is impossible; protocols resorting to
additional degrees of freedom have been proposed. Our analysis shows
that HBSA enables the teleportation of an arbitrary state encoded in
both polarization- and spatial-mode with a  success probability of
100\%.

\begin{figure}[!ht]
\begin{center}
\includegraphics[width=4.5cm,angle=0]{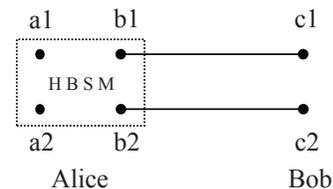}
\caption{The procedure of teleportation of an unknown single
particle state in the polarization- and spatial-mode degrees of
freedom with a hyperentanglement-state channel, resorting to HBSA.}
\end{center}
\end{figure}

Suppose that the photon $A$ teleported in Alice's laboratory is in
an arbitrary state $|\varphi\rangle_{A}$ in both the polarization
and   spatial-mode degrees of freedom, that is,
\begin{eqnarray}
|\varphi\rangle_{A}=(\alpha|H\rangle+\beta|V\rangle)\otimes(\gamma|a1\rangle+\delta|a2\rangle),
\end{eqnarray}
where $|\alpha|^{2}+|\beta|^{2}=1$ and
$|\gamma|^{2}+|\delta|^{2}=1$. Alice and Bob share a
hyperentanglement-state channel $BC$ with the form
\begin{eqnarray}
|\Phi^{+}\rangle_{BC}=\frac{1}{2}(|HH\rangle+|VV\rangle)\otimes(|b1c1\rangle+|b2c2\rangle).
\end{eqnarray}
The principle of teleportation of an unknown single-particle state
in two degrees of freedom with hyerentanglement is shown in Fig.5.
If Alice performs a HBSA on   photons $A$ and $B$, the whole system
will evolves as
\begin{eqnarray}
&&|\varphi\rangle_{A}\otimes|\Phi^{+}\rangle_{BC}\nonumber\\
&=& \frac{1}{2}(\alpha|HHH\rangle+\alpha|HVV\rangle+\beta|VHH\rangle\nonumber\\
&& \; + \beta|VVV\rangle)_{ABC}
\otimes(\gamma|a1b1c1\rangle+\gamma|a1b2c2\rangle\nonumber\\
&& \; + \delta|a2b1c1\rangle+\delta|a2b2c2\rangle)_{ABC}\nonumber\\
&=& \frac{1}{4}[(|\phi^{+}\rangle_{P}(\alpha|H\rangle+\beta|V\rangle)+ |\phi^{-}\rangle_{P}(\alpha|H\rangle
-\beta|V\rangle))\nonumber\\
&& \;\; + |\psi^{+}\rangle_{P}(\alpha|V\rangle+\beta|H\rangle) + |\psi^{-}\rangle_{P}(\alpha|V\rangle
-\beta|H\rangle)]_{ABC}\nonumber\\
&& \; \otimes [|\phi^{+}\rangle_{S}(\gamma|c1\rangle+\delta|c2\rangle) + |\phi^{-}\rangle_{S}(\gamma|c1\rangle
-\delta|c2\rangle)\nonumber\\
&& \;\; + |\psi^{+}\rangle_{S}(\gamma|c2\rangle+\delta|c1\rangle) +
|\psi^{-}\rangle_{S}(\gamma|c2\rangle-\delta|c1\rangle)]_{ABC}.\nonumber\\\label{Tevoluation}
\end{eqnarray}
If Alice obtains the outcomes $\vert \phi^\pm\rangle_P\vert
\phi^\pm\rangle_S$, $\vert \phi^\pm\rangle_P\vert
\psi^\pm\rangle_S$, $\vert \psi^\pm\rangle_P\vert
\phi^\pm\rangle_S$, or $\vert \psi^\pm\rangle_P\vert
\psi^\pm\rangle_S$, the photon in Bob's hand will be in the states
$(\alpha|H\rangle \pm \beta|V\rangle)_P(\gamma|c1\rangle \pm
\delta|c2\rangle)_S$, $(\alpha|H\rangle \pm
\beta|V\rangle)_P(\gamma|c2\rangle \pm \delta|c1\rangle)_S$,
$(\alpha|V\rangle \pm \beta|H\rangle)_P(\gamma|c1\rangle \pm
\delta|c2\rangle)_S$, or $(\alpha|V\rangle \pm
\beta|H\rangle)_P(\gamma|c2\rangle \pm \delta|c1\rangle)_S$,
respectively. With the results published by Alice, Bob can recover
the unknown state $|\varphi\rangle_{A}$ with two local unitary
operations on his photon $C$. For example, if Bob obtains the state
$(\alpha|V\rangle - \beta|H\rangle)_P(\gamma|c2\rangle -
\delta|c1\rangle)_S$, he first performs a unitary operation
$-i\sigma_y \equiv \vert H\rangle\langle V\vert - \vert
V\rangle\langle H\vert$ on the photon $B$ in the polarization degree
of freedom (each of the two spatial modes $c1$ and $c2$) and then
introduces a relative phase $\pi$ in the spatial mode $c1$ which can
be accomplished with a $\lambda/2$ wave plate. With the exchange of
the two spatial modes $c1$ and $c2$, Bob can obtain the unknown
single-particle state in two degrees of freedom
$|\varphi\rangle_{B}=(\alpha|H\rangle+\beta|V\rangle)\otimes(\gamma|c1\rangle+\delta|c2\rangle)$.
The other cases are similar to this one with or without a little
modification.

To date, quantum teleportation protocols have been demonstrated with
a success probability of 50\%, resorting to linear optical
Bell-state measurements, such as two-photon interference,
single-photon detection, and polarization analysis. Walborn \emph{et
al.} \cite{hyper2} also discussed quantum teleportation protocol
with hyperentanglement in 2003. However, in their protocol, they can
teleport an arbitrary state encoded in either the polarization or
the momentum degree of freedom with a success probability of  50\%,
that is, the same probability as the protocol with two-photon
polarization BSA based on linear optics. Compared with the
conventional polarization teleprotation protocols, it does not offer
more advantages. The present teleportation scheme with
 hyperentanglement provides us a way of teleporting a quantum state in both   polarization
 and spatial-mode degrees of freedom. Compared with the
conventional teleportation protocols, more quantum information can
be transmitted.

\subsection{Entanglement swapping  with hyperentangled states}

Another interesting application of HBSA is quantum entanglement
swapping, which enables one to entangle two quantum systems that
have never being interacted with each other
\cite{swapping1,swapping2}. Entanglement swapping has been widely
applied in quantum repeaters
\cite{repeater1,repeater2,DLCZ,zhao,hybid1,hybid2,swappingexperiment1,swappingexperiment2}.
In a practical transmission for long-distance quantum communication,
the photon losses increase exponentially with the length of the
communication channel \cite{repeater1}. In order to overcome the
photon losses, the whole transmission channel is usually divided
into many segments and the length of each segment is comparable to
the channel attenuation length. Entanglement is  first generated in
each segment and then extended to a greater length by connecting two
adjacent segments with entanglement swapping.

The existing entanglement-swapping protocols usually focused on  the
Bell states with only one degree of freedom, that is, the
polarization of photons. Here we  show that we can also perform the
entanglement swapping with hyperentangled states, which will improve
largely the channel capacity in long-distance quantum communication.

\begin{figure}[!ht]
\begin{center}
\includegraphics[width=6cm,angle=0]{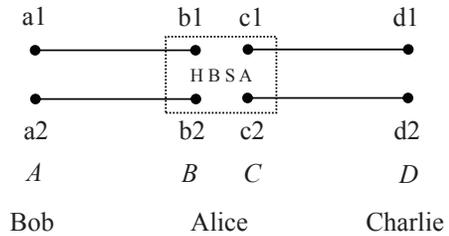}
\caption{Schematic diagram of hyperentanglement swapping in the
polarization and  spatial-mode degrees of freedom. The initial
hyperentangled states are prepared in nodes $AB$ and $CD$ (also the
four photons). After Alice performs the HBSA on the two photons
$BC$, Bob and Charlie can get the hyperentangled state between nodes
$A$ and $D$. $a1b1$ ($a2b2$) are the different spatial modes for
each hyperentangled state.}
\end{center}
\end{figure}

Let pairs $AB$ and $CD$ be in the following hyperentangled states:
\begin{eqnarray}
|\Phi^{+}\rangle_{AB}&=&|\phi^{+}\rangle^{AB}_{P}\otimes|\phi^{+}\rangle^{AB}_{S}\nonumber\\
&=&\frac{1}{2}(|HH\rangle+|VV\rangle)\otimes(|a1b1\rangle+|a2b2\rangle),\\
|\Phi^{+}\rangle_{CD}&=&  |\phi^{+}\rangle^{CD}_{P}\otimes|\phi^{+}\rangle^{CD}_{S}\nonumber\\
&=&\frac{1}{2}(|HH\rangle+|VV\rangle)\otimes(|c1d1\rangle+|c2d2\rangle).
\end{eqnarray}
The subscript $P$ denotes the polarization part of the
hyperentangled state and  $S$ is the spatial-mode part. The
superscripts $A$ and $B$ denote that the particles are in nodes $A$
and $B$, respectively, as shown in Fig. 6. That is, Alice shares a
photon pair $AB$ with Bob. Also, she shares a photon pair $CD$ with
Charlie. The task of this entanglement swapping protocol is to
entangle the two photons $A$ and $D$ in both   polarization and
 spatial-mode degrees of freedom.

For entanglement swapping of a hyperentangled state, Alice performs
HBSA on the two particles $B$ and $C$, as shown in Fig. 6. The whole
system evolves as:
\begin{eqnarray}
&&|\Phi^{+}\rangle_{AB}\otimes|\Phi^{+}\rangle_{CD}\nonumber\\
&=&\frac{1}{4}(|HHHH\rangle+|HHVV\rangle
+|VVHH\rangle+|VVVV\rangle)\nonumber\\
&\otimes&(|a1b1c1d1\rangle+|a1b1c2d2\rangle
+|a2b2c1d1\rangle+|a2b2c2d2\rangle)\nonumber\\
&=&\frac{1}{4}(|HHHH\rangle+|HVHV\rangle
+|VHVH\rangle+|VVVV\rangle)\nonumber\\
&\otimes&(|a1d1b1c1\rangle+|a1d2b1c2\rangle+|a2d1b2c1\rangle+|a2d2b2c2\rangle)\nonumber\\
&=&\frac{1}{4}[(|\phi^{+}\rangle^{AD}_{P}|\phi^{+}\rangle^{BC}_{P}+|\phi^{-}\rangle^{AD}_{P}|\phi^{-}\rangle^{BC}_{P}\nonumber\\
&& \;\; +|\psi^{+}\rangle^{AD}_{P}|\psi^{+}\rangle^{BC}_{P}+|\psi^{-}\rangle^{AD}_{P}|\psi^{-}\rangle^{BC}_{P})\nonumber\\
&& \;\; \otimes(|\phi^{+}\rangle^{AD}_{S}|\phi^{+}\rangle^{BC}_{S}+|\phi^{-}\rangle^{AD}_{S}|\phi^{-}\rangle^{BC}_{S}\nonumber\\
&& \;\;
+|\psi^{+}\rangle^{AD}_{S}|\psi^{+}\rangle^{BC}_{S}+|\psi^{-}\rangle^{AD}_{S}|\psi^{-}\rangle^{BC}_{S})].\label{entangleswapping}
\end{eqnarray}
If Alice obtains the outcome $|\Phi^{+}\rangle_{BC}=\vert
\phi^+\rangle^{BC}_P\vert \phi^+\rangle^{BC}_S$, the two photons
located in  nodes $A$ and $D$ will be in the hyperentangled state
$|\Phi^{+}\rangle_{AD}=\vert \phi^+\rangle^{AD}_P\vert
\phi^+\rangle^{AD}_S$. The outcomes will lead to the other
hyperentangled states, such as $\vert \phi^+\rangle^{AD}_P\vert
\phi^-\rangle^{AD}_S$, $\vert \phi^-\rangle^{AD}_P\vert
\phi^\pm\rangle^{AD}_S$, $\vert \phi^-\rangle^{AD}_P\vert
\psi^\pm\rangle^{AD}_S$, $\vert \psi^\pm\rangle^{AD}_P\vert
\phi^\pm\rangle^{AD}_S$, and $\vert \psi^\pm\rangle^{AD}_P\vert
\psi^\pm\rangle^{AD}_S$. Moreover, it is, in principle, not
difficult for Bob and Charlie to transform their hyperentangled
states into the form $|\Phi^{+}\rangle_{AD}$. For instance, if Bob
and Charlie obtain the state $\vert \psi^-\rangle^{AD}_P\vert
\psi^-\rangle^{AD}_S=\frac{1}{2}(\vert HV\rangle - \vert
VH\rangle)(\vert a1d2\rangle - \vert a2d1\rangle)$, they can obtain
the state $|\Phi^{+}\rangle_{AD}=\frac{1}{2}(\vert HH\rangle + \vert
VV\rangle)(\vert a1d1\rangle + \vert a2d2\rangle)$ in the way that
Charlie performs an
 operation $-i\sigma_y$ in polarization (the two spatial modes, i.e., the two paths $d1$ and $d2$) and then exchanges
 the two spatial modes after he introduces a phase
$\pi$ in the spatial mode $d1$ with a $\lambda/2$ wave plate.

In essence, the present hyperentanglement-swapping protocol can be
divided into two processes, that is the entangled swapping for
entanglement states in polarization and that in spatial modes. Each
entanglement-swapping process is independent of the other.
Meanwhile, they must be swapped simultaneously. If we only perform
the Bell-state measurement on photons $B$ and $C$ in the
polarization degree of freedom, photons $A$ and $D$ will be
entangled in the polarization degree of freedom but their state in
spatial modes may be a mixed one..

\section{discussion and summary}

A hyperentangled BSA is far different from a
hyperentanglement-assisted BSA as the latter is used to only analyze
the  Bell states in the polarization degree of freedom and the other
degree of freedom, such as momentum and time-bin, is an additional
auxiliary system and is consumed in the analysis. This HBSA protocol
shows that with cross-Kerr nonlinearity, one can also perform a
complete HBSA. During a HBSA process, the entanglements in different
degrees of freedom can be manipulated independently. The key element
in this scheme is QND as it provides us the way to manipulate the
photons but not destroy them. This protocol also reveals that
complete HBSA with only linear optical elements is also impossible.

In the process of describing the principle of our HBSA scheme, we
mainly exploit the cross-Kerr nonlinearity to construct the
parity-check gate.  We should acknowledge that although a lot of
works have been studied on cross-Kerr nonlinearity \cite{RMP}, a
clean cross-Kerr nonlinearity in the optical single-photon regime is
still quite a controversial assumption, especially with current
technology. In Ref. \cite{kok02}, Kok \emph{et al.} showed that
operating in the optical single-photon regime, the Kerr phase shift
is only $\tau \approx 10^{-18}$. With electromagnetically induced
transparent materials, cross-Kerr nonlinearities of $\tau
\thickapprox  10^{-5}$ can be obtained. As pointed out by
Gea-Banacloche \cite{julio} recently, the large phase shifts via the
giant Kerr effect with single-photon wave packets is impossible at
present. These results also agree with the previous works by Shapiro
and Razavi \cite{shapiro1,shapiro2}. The weak cross-Kerr
nonlinearity will make the phase shifts $\theta_1 + \theta_3$,
$\theta_2 + \theta_4$, $\theta_1 + \theta_4$, and $\theta_2 +
\theta_3$  of the coherent state become extremely small, which will
be hard to detect. That is to say, using homodyne detector, it is
difficult to determine the phase shift due to the impossible
discrimination of two overlapping coherent states, which will
decrease the success probability of the present HBSA scheme. In
2003, Hofmann \emph{et al.} \cite{Hofmann} showed that a phase shift
of $\pi$ can be achieved with a single two-level atom in a one-sided
cavity. In 2010, Wittmann \emph{et al.} \cite{wittmann} investigated
quantum measurement strategies capable of discriminating two
coherent states using a homodyne detector and a
photon-number-resolving (PNR) detector. In order to lower the error
probability, the postselection strategy is applied to the
measurement data of homodyne detector as well as a PNR detector.
They showed that the performance of the new displacement-controlled
PNR is better than that of a homodyne receiver. That is, the present
HBSA scheme may be feasible if we choose a suitable Kerr nonlinear
media and some good quantum measurement strategies on coherent
beams. Moreover, quantum gates based on optical nonlinearities have
attracted a lot of attention
\cite{kok,friedler1,friedler2,friedler3} in recent years, which will
impel the development of optical nonlinearity techniques.

In fact, the present HBSA scheme requires that the positive and
negative signs in the phase shifts of the coherent beam can not be
distinguished. That is, it is unnecessary for us to have
interaction-induced phases of both positive and negative sign in the
present scheme. Here a cross-Kerr nonlinearity in QNDs is only used
to make a parity check for two photons and  other elements can also
be used to construct QNDs \cite{e1,e2,e3} for this HBSA scheme.

In summary, we have proposed a complete HBSA scheme with cross-Kerr
nonlinearity. We use the cross-Kerr nonlinearity to construct
parity-check measurements and analyze Bell states in different
degrees of freedom of photons. We also discussed its applications in
quantum teleportation and entanglement swapping in two different
degrees of freedom simultaneously. We concluded that one can
teleport a particle in more than one degree of freedom if a
hyperentanglement channel is set up and a hyperentanglement
Bell-state measurement is permitted perfectly. We also revealed that
quantum communication based on hyperentanglement is possible as we
can set up a hyperentanglement quantum channel for a long-distance
quantum communication with hyperentanglement swapping. All these
results may be useful in practical applications in quantum
information.

\section*{ACKNOWLEDGEMENTS}

This work is supported by the National Natural Science Foundation of
China under Grant Nos. 10775076, 10874098, and 10974020, the
National Basic Research Program of China (Grant Nos. 2006CB921106
and 2009CB929402), China Postdoctoral Science Foundation under Grant
No. 20090460365, and the Specialized Research Fund for the Doctoral
Program of Education Ministry of China (Grant No. 20060003048).

\end{document}